# Noncovalent Functionalization of Boron Nitride Nanotubes in Aqueous Media

*Zhenghong Gao,[†*] Chunyi Zhi,[‡] Dimtri Golberg,[‡] Yoshio Bando,[‡] Takeshi Serizawa[§]*

[†]*Laboratoire Photonique, Numérique et Nanosciences (LP2N), Institut d'Optique Graduate School, CNRS & Université de Bordeaux, 351 cours de la libération, Talence Cedex 33405, France.*

[‡]*International Center for Materials Nanoarchitectonics (MANA), National Institute for Materials Science (NIMS), Namiki 1-1, Tsukuba, Ibaraki 305-0044, Japan.*

[§]*Department of Organic and Polymeric Materials, Tokyo Institute of Technology, 2-12-1-H121 Ookayama, Meguro-ku, Tokyo 152-8550, Japan.*

**ABSTRACT.** Boron nitride nanotubes (BNNTs) are of intense scientific interests due to their unique physiochemical properties and prospective applications in various nanotechnologies. A critical problem hampering the application processing of BNNTs is the outer sidewall functionalization, which is primarily acquired to lead BNNTs dispersible in various solvents. Furthermore, the surface of BNNTs should be intelligently designed and precisely controlled to satisfy the specific demands of different applications. For these purposes, covalent and noncovalent approaches have been factually developed for opening up the key door of applications. Importantly, wrapping the outermost sidewall of BNNTs with either water-soluble polymers or biomolecules through weak noncovalent interactions has been proved to be efficient for giving BNNTs considerable dispersity in aqueous media, and endowing novel chemical functions to the BNNTs with almost no change in their pristine physiochemical properties. This contribution is made to summarize recent progresses, and further addresses the future perspectives on the noncovalent functionalization of BNNTs for promoting their application processing.

## 1.1 Introduction

In the recent years, one-dimensional (1-D) nanomaterials[1] such as nanotubes,[2] nanowires,[3] and nanorods[4] have attracted intensively scientific interests because of their unique nanoscale size and 1-D geometry-related physiochemical properties.[5] The properties of 1-D nanomaterials are closely related to their new quantum mechanical effect, size effect, and surface effect; they can be far different from those for convensional bulk materials.[1] In turn, the rising of 1-D nanomaterials produces great promises for developing novel nanosystems towards many urgent technologies such as high-efficiency energy conversion devices[6] and environmentally friendly technologies.[7] Carbon nanotubes (CNTs) are the representatives of 1-D nanomaterials. The scientific interests to CNTs are greatly proved by their unique 1-D nanostructure and therefore multiple outstanding properties of electronic, optic, thermal, and mechanic ones.[8] Structurally, this 1-D carbon allotrope is made by rolling up a two-dimensional (2-D) graphene sheet-a one-carbon atom thick layer along a certain direction. The roll-up direction is theoretically represented by a pair of defined indices (n, m).[9] The integer n and m denote the number of the unit vectors along two directions in the honeycomb crystal lattice of graphene. The CNTs are classified into armchair (*n, n*), zigzag (*n, 0*), and chiral (*n, m*) properties, depending on the different roll-up directions.[9] On the other hand, CNTs also be classified into single-walled CNTs (SWCNTs), double-walled CNTs (DWCNTs), and multi-walled CNTs (MWCNTs), according to the graphene sheet numbers. The properties of those CNTs are thereby quite different from each other.[10] Even though more than 20 years have already been passed since the clear growing and observation of the tubular structure of CNTs,[2] even the first report on this filament form of carbon materials was factually much earlier.[11] To date, the applications of CNTs are still hampered by the pristine indispersity in most aqueous and organic solvents. In other words, as-produced CNTs closely aggregate as bundles through the strong van der Waals interactions between the tubes' sidewalls.[12] At the same time, the electrical properties of SWCNTs are correlated with their specific chirality.[13] SWCNTs can thereby be metallic or semiconducting depending on different chirality. Although several methods have been reported being efficient for the isolation of CNTs with specific chirality, unfortunately, almost all those methods are currently limited in a batch. The efficient methods for the precise isolation of SWCNTs with specific chirality in a large scale to fit the needs of industry applications are still much merely. In this sense, the functionalization of the sidewall of CNTs are primarily acquired not only to make them dispersible in aqueous and organic solvents, but also for extracting samples rich in single chiral tubes. Noncovalent[14] and covalent[15,16] approaches have been explored to solve this problem so far.

Boron nitride nanotubes (BNNTs)[17] are the representative inorganic analogues of CNTs (**Figure 1**).[18] The theoretical prediction of BNNTs[19,20] and the experimental synthesis[17] have been successfully reported in 1994 and 1995, respectively. BNNTs are structurally similar to CNTs except that B and N atoms alternatively substitute C atoms in the graphene-like honeycomb crystal lattice, giving a one-atom thick hexagonal BN layer.[21] As shown in **Figure 2**, the same indices (n, m) are applicable for definition of the chirality of these tubes as those for CNTs.[22,23] However, in contrast to CNTs, BNNTs have an electric band-gap around 5.5 eV independent of structural chirality and diameter.[20,24] The distinct wide electric band-gap and high chemical stability in high temperature/oxidation environments[25] allow BNNTs invaluable for high-performance nanocomposite materials[26,27] and biological application such as drug delivery.[28,29] Similar to CNTs, BNNTs hold highly hydrophobic sidewalls, and tightly aggregate together with their adjacent tubes via van der Waals forces,[30] resulting that as-produced BNNTs exist as bundles or ropes. Therefore, BNNTs are un-dispersible in many aqueous and organic solvents.[31] The potential applications of BNNTs are hampered by their undispersity, similarly to those for CNTs. It is strongly required to clear this obstacle by exploiting new functionalization approaches by modifying the hydrophobic properties of BNNTs' sidewalls to make them primarily dispersible in many solvents, particularly in aqueous solutions.[32] The functionalizations of the sidewalls of CNTs with various synthetic polymers via covalent grafting or noncovalent wrapping have been widely investigated mostly due to the rapid progress in the synthesis methods. However, the corresponding studies on BNNTs were lied far behind because of the serious lack of an efficient method for producing BNNTs with high quality and purity.[32] Several years ago, a key breakthrough for synthesizing BNNTs in a gram scale has been achieved by a pioneer group in the BNNTs research at National Institute for Materials Science -NIMS in Japan.[33-36] Since then, the research activities on BNNTs started to grow very quickly. So far, many methods have been exploited including low temperature growth[37] and patterned formation on substrates.[38] However, the synthesis and post-purification of BNNTs are factually much difficult in comparison to the same issues for CNTs.[34] Most samples currently involved in the research are still multi-walled BNNTs (MWBNNTs). High quality single-walled BNNTs (SWBNNTs) sample are relatively rare in both laboratory and commercial market. Although the size and properties of SWBNNTs and MWBNNTs are different from each other, their basic structures of the outermost sidewalls are commonly the same. Therefore, the chemical functionalization approaches for modifying MWBNNTs should also be alternatives for SWBNNTs.

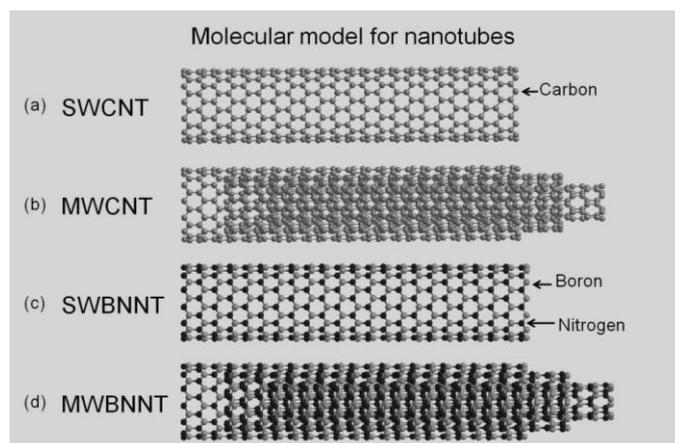

**Figure 1.** Molecular model of (a) SWCNT, (b) MWCNT, (c) SWBNNT and (d) MWBNNT.

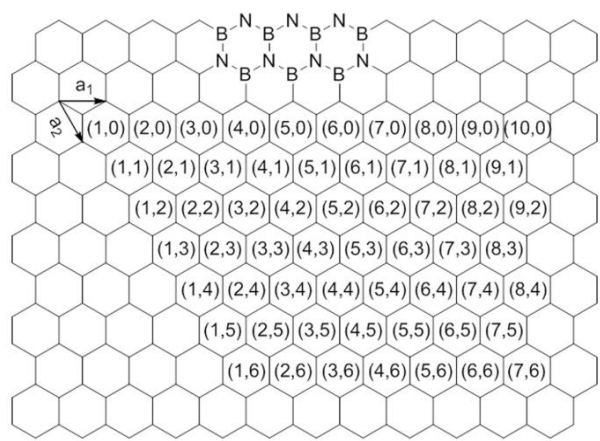

**Figure 2.** Vectors (n, m) for single-walled BNNT on an h-BN sheet.

## 1.2 Overview of Functionalization Approaches for BNNTs

For the sidewall functionalization of BNNTs, there are generally four main approaches available. The first functionalization approach is to covalently attach alien chemical groups to the -NH and/or -NH$_2$ defect sites presented on the sidewall of BNNTs, which are induced during the imperfect crystalline growth process or by chemical post-treatments such as harsh acid oxidation. Another covalent approach is to form chemical bonds by the use of the boron sites based on boron chemistry. For examples, Zhi *et al.* have developed a covalent approach based on the reaction between the -COCl group of stearoyl chloride and the amino groups (defect sites) on the BNNTs after refluxing the mixture for 120 h at 100 ℃.[31] The resultant BNNTs were dispersible in organic solvents such as chloroform, *N*,*N*-dimethylacetamide, tetrahydofuran, *N*,*N*-dimethylformamide, acetone, toluene, and ethanol. Other covalent approaches for the functionalization of BNNTs include modification with amine-terminated

poly(ethylene glycol) which forms ionic bonds with B sites on BNNTs sidewalls by heating 100 °C for several hours;[30] peeling the B-N bond by cyclic treating BNNTs with dimethyl sulfoxide (DMSO), in which DMSO molecules were added to the sidewalls of BNNTs through nucleophilic attack of O on B and electrophilic attack of S on N under the mixing processes of sonication and heating at 180 °C for 6 h;[39] fluorination by inducing F atoms during the stage of BNNTs growth, here F were covalently doped to N atoms due to their close electronegativities;[40] and interaction with Lewis bases under heating at 70 °C for 12 h and followed by sonicating for a few minutes, in which B on BNNTs surface behaved as Lewis acids and interacted with various Lewis bases.[41] The advantage of a covalent way is that the formed chemical bonds are stable and durable; the main drawbacks of a covalent approach are the changes of the pristine crystalline structure and properties, in particular, the destruction of outer surface/sidewall structure, even leading to a poor biocompatibility.

On the other hand, attachment of polymers and biomolecules onto BNNTs through noncovalent interactions such as π-π stacking interactions provides a novel and gentle approach for functionalization. For example, it has been demonstrated that a conjugated polymer of poly[*m*-phenylenevinylene-co-(2,5-dioctoxy-*p*-phenylenevinylene)] (PmPV) is able to noncovalently wrap BNNTs.[42] PmPV-wrapped BNNTs were dispersible in chloroform, *N*,*N*-dimethylacetamide, and tetrahydofuran, but they were un-dispersible in water and ethanol. It is easily notable that many approaches are workable for making BNNTs dispersible in organic solvents, but still un-dispersible in aqueous phases. However, the aqueous dispersion is preferable by most of biological applications which are the major prospective of BNNTs, but only few efforts have been addressed to disperse BNNTs in aqueous phases. For example, Zhi *et al.* reported the dispersion of BNNTs in water using a single-strand DNA wrapping approach via π-π stacking interactions.[43] Chen *et al.* reported the functionalization of BNNTs using an amphipathic dendritic structure comprising synthetic carbohydrate ligands at the chain ends that enable specific binding to receptors in solution. In this case, a pyrene group at the dendrimer focal point allows that these dendrimers are able to interact with the isoelectronic BNNT surface through π-π stacking and hydrophobic interactions. The simple noncovalent absorption of amphipathic dendritic structures enables the BNNTs surface to displaying the glycodendrimers capable of interacting with proteins and cells. The method should facilitate applications of BNNTs in biosensing and bioimaging.[44] Wang *et al.* reported the aqueous noncovalent functionalization of BNNTs using an anionic perylene derivative, namely perylene-3,4,9,10-tetracarboxylic acids tetrapotassium salts (PTAS).[45] The interactions were the π-π stacking interactions between PTAS and BNNTs.[46] A plenty of surface-attached COO⁻ groups gave in turn the dispersity and chemical activity to BNNTs. PTAS-functionalized BNNTs can superiorly bind to a large variety of metal ions. Importantly, by employing a high temperature vacuum heating of the BNNTs to 1180 ℃ in a rate of 2 ℃/min and continuous isothermal annealing the samples at the same temperature over 2 h, PTAS-BNNTs were converted into C-doping BNNTs which exhibited a considerable semiconducting properties. In addition, the filling of the inner tube channel with small molecules[47] or inorganic nanoparticles[48-51] is another approach for functionalizing BNNTs.

In sum up of the above-mentioned advancements in the dispersion and functionalization of BNNTs, although notable covalent and noncovalent approaches have been developed, many new approaches are still highly desired for extending their processabilities of various technical applications.

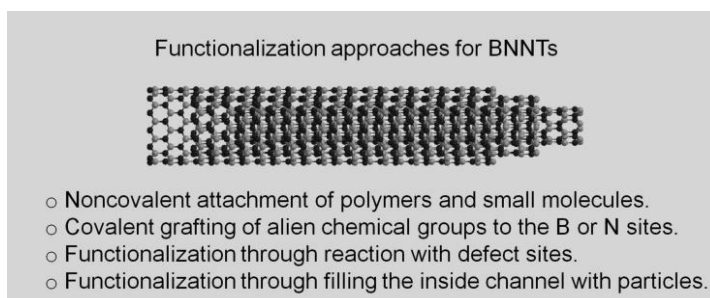

**Figure 3.** Functionalization approaches for BNNTs.

## 1.3 Noncovalent Approaches for Dispersion and Functionalization of BNNTs

Among the various approaches for surface functionalization of both CNTs and BNNTs, noncovalent attachment of polymers[52,53] or small molecules[54,55] based on the weak inter-molecular interactions, such as π-π stacking,[55] cation-π interactions,[56] and hydrogen bonds,[57] has been realized preservation of the pristine electric and optical properties which are particularly essential for biological applications.[14] On the other hand, due to the diversity of polymers and small molecules, different alien chemical groups can be intensively attached onto the sidewall of NTs to endow different chemical functions depending on the occasions of applications. As well, noncovalently attached molecules are possible to be removed from the sidewalls by suitable methods with minimum interruption to the intrinsic properties, which allow the recovery of the nanotube surface.[14] In addition, noncovalent polymer wrapping is efficient for removal of the catalysis species, purification,[58] and selective dispersion[59] of nanotube samples via a simple dispersion and extraction processes. In this sense, the development of approaches for noncovalent functionalization of BNNTs is increasing scientific interests.

### 1.3.1 Aim and Prospects

The unique 1-D geometry and outstanding chemical stability of BNNTs promoted them as promising candidates for composite materials[27,60-63] (mainly insulated thermal conducting polymeric composite) and biological applications,[44,64] such as drug delivery system (DDS),[29] magnetic resonance imaging (MRI) contrast agents,[65] boron neutron capture therapy,[66] and cancer cell treatments.[67] The strong van der Waals interactions between BNNTs makes them existing as the bundles in their as-produced state. This results into the pristine undispersity of BNNTs in common aqueous and organic solvents.[32] However, for many applications, particularly bio-related fields, dispersion in water or physiological buffer is preferred primarily. On the other hand, new biological functions are acquired importantly to BNNTs for increasing the availability in biological fields. Also, the biocompatibility of BNNTs needs to be improved in most cases of biological applications.[68] The noncovalent approaches in aqueous solutions are efficient for functionalizing the sidewalls of BNNTs to address a broad range of applications. In this content, the contribution is made to give a comprehensive understanding on the subject of noncovalent functionalization of BNNTs with either synthetic polymers or biomolecules in aqueous phases.

### 1.3.2 Noncovalent Functionalization of BNNTs with Synthetic Polymers

Synthetic polymers containing aromatic subunits can interact with the sidewall of BNNTs via π-π stacking interactions.[69] The earliest report on the wrapping BNNTs with a π-conjugate conducting polymer-PmPV was illuminated by Zhi *et al*.[42] The functionalization was conducted by simply sonicating the mixture of BNNTs with PmPV. A homogeneous dispersion was obtained after removing insoluble materials via centrifugation process. The resultant PmPV-wrapped BNNTs were dispersible in organic solvents, including chloroform, *N,N*-dimethylacetamide, tetrahydrofuran, etc., whereas they were un-dispersible in water, ethanol, etc. Velayudham *et al*. investigated the noncovalent functionalization of BNNTs using three conjugated poly(*p*-phenylene ethynylene)s (PPEs) (Polymers A and B) and polythiophene (Polymer C) derivatives through the strong π-π stacking interactions between the polymers and BNNTs.[70] The chemical structures of these polymers are shown in **Figure 4**.

The resultant BNNTs are dispersible in organic solvents such as chloroform, methylene chloride, and tetrahydrofuran. A layer of the PPE derivative can be clearly seen on the surface of BNNTs in the high-resolution transmission electron microscopic (HR-TEM) image (**Figure 5**). The absorption and emission of the PPE derivatives attaching on BNNTs showed red shifts in comparison with free PPEs due to the enhanced planarization, while the polythiophene derivative showed blue shifts by the strong disruption of its π plane.

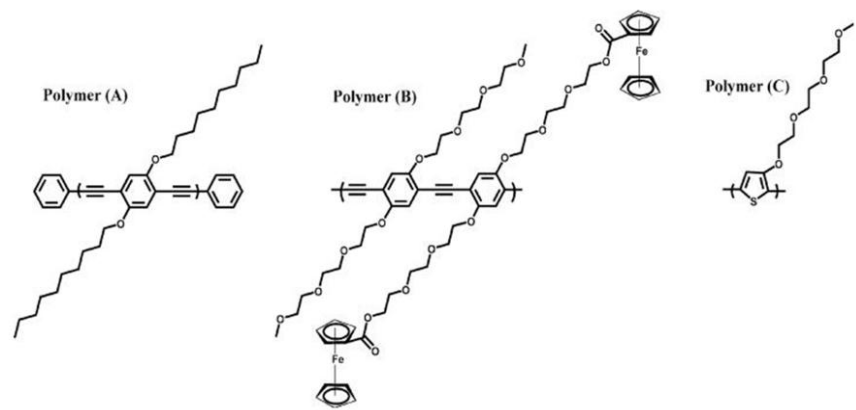

**Figure 4.** Chemical structures of the PPE derivatives (Polymers A and B) and the regioregular head-to-tail polythiophene derivative (Polymer C). (Reproduced with permission from ref 70. Copyright (2010) American Chemical Society).

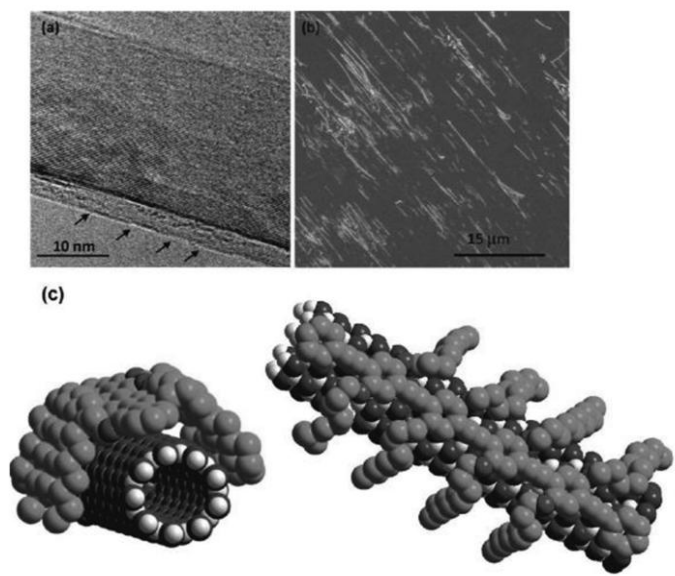

**Figure 5.** (a) HR-TEM image at the BNNT sidewalls functionalized with the PPE derivative (Polymer A). A layer of the PPE derivative can be clearly seen as indicated by arrows. (b) SEM image of aligned BNNTs functionalized with Polymer A. (c) Schematic models of composite material of BNNTs and Polymer A at various viewing angles. (Reproduced with permission from ref 70. Copyright (2010) American Chemical Society).

Even though it has been shown that several examples of polymers can function as dispersants for functionalizing the sidewall of BNNTs, the details on the interactions of different polymers with BNNTs have not yet been fully understood.[71] Our group has recently investigated the potential of different water-soluble polymers, including a poly(*p*-phenylene) derivative ((-)PPP), poly(xylydiene tetrahydrothiophene) (PXT), poly(sodium styrene sulfonate) (PSS), poly(sodium vinyl sulfonate) (PVS), and poly(sodium acrylate) (PAA), for dispersing BNNTs in water (**Figure 6**).[72] Our investigations have shown that (-)PPP has the highest potential for dispersing BNNTs among the polymers. This result indicated that polymers with an extensive π-plane show stronger interactions with BNNTs than those with a smaller π-plane, suggesting that the π-π stacking interactions play a very important role in the functionalization of BNNTs with polymers. Importantly, by simple chemical conversion of PXT to PPV, we have obtained a film composed of PPV-functionalized BNNTs. Contact angle (CA) measurements suggested that this film exhibited a CA of 151 ± 1 °which permits it non-wetting/superhydrophobic (**Figure 7**). This finding is promising for the construction of materials with unfouling surfaces. A molecular dynamic (MD) simulation study on the interfacial binding interactions of BNNTs and various polymers including PmPV, polystyrene (PS), and polythiophene (PT) using a density functional theory (DFT) has been carried out by Nasrabadi *et al.* in 2010.[71] Computer calculations of dihedral angle (θ) were conducted to figure out the interaction energy between BNNTs and polymer molecules, and the morphology of polymers stacked onto the BNNT surface. The results pointed out that the specific monomer structure of polymer and BNNT radius had strong influence on the interaction strength, but the influence of temperature is likely negligible. Among PmPV, PS and PT, the order for stronger binding was PT, PmPV, and PS. Moreover, they have revealed that the BNNT-polymer interactions are much stronger that those of the similar CNT-polymer composites, supporting that those BNNTs are excellent candidates for the construction of high-performance polymer composite materials.

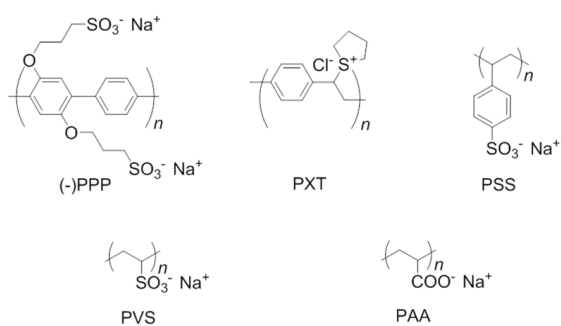

**Figure 6.** Chemical structures of water-soluble several synthetic polymers. (Reproduced with permission from ref 72. Copyright (2012) Nature Publishing Group).

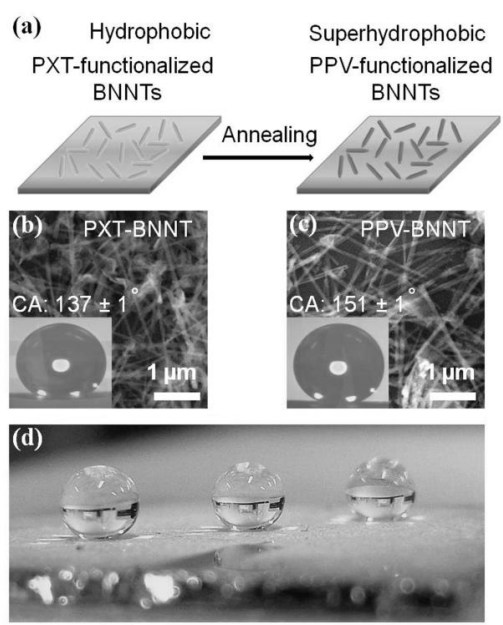

**Figure 7.** (a) Schematic image of the conversion of PXT into PPV on the BNNT sidewalls via thermal treatment, Scanning electron microscopy (SEM) image of the film surfaces coated with (b) PXT- and (c) PPV-functionalized BNNTs coated on a Si wafer, and (d) an optical photo of water droplets on a superhydrophobic PPV-functionalized BNNTs. The insets in (b) and (c) show water droplets used for contact angle measurements. (Reproduced with permission from ref 72. Copyright (2012) Nature Publishing Group).

### 1.3.3 Noncovalent Functionalization of BNNTs with Biomolecules

Even though many synthetic polymers are available for functionalizing BNNTs sidewall, most of synthetic polymers are supplied by organic and polymer synthesis, which is sometimes complicated. Another problem is that the biocompatibilities of synthetic polymers are still almost secrets. On this standpoint, water-soluble biomolecules offer great opportunities for the functionalization of BNNT due to their diversity in structures and functions.[73] They are low-cost, biocompatible, and no need for precise synthesis. Biomolecules refer to chemical compounds found in living organisms. The most prominent classes of biomolecules are peptides, proteins, DNAs, RNAs, saccharides, and lipids. Biomolecules often show specific complementary interactions such as antibody-antigen, hormone-receptor, and protein-protein/nucleic acid-nucleic acid/nucleic acid-protein interactions. In combination with biomolecules' important natures of self-assembly and biological evolution, functionalization of nanomaterials with biomolecules appears as an important research area of materials' chemistry.[74] It has been well proven that many biomolecules can be employed as dispersants and/or functional reagents for dispersing and functionalizing CNTs.[74] With consideration of the similarity in the structure, many biomolecules can be similarly efficient for dispersing and functionalizing BNNTs. The research on this topic has just started to draw full attention recently.

In our efforts, taking many advantages of biomolecules into account, dispersing and functionalizing BNNTs in an aqueous solution was performed using peptides,[75] flavin mononucleotide (FMN),[76] DNA[43]/nucleotides[77] and polysaccharides.[78] These studies were firstly aimed at making BNNTs well dispersible in an aqueous solution via biomolecules functionalization, thereby opening a new pathway for bio-related applications. The second objective of these studies was to endow new physiochemical properties to BNNTs after functionalizing with biomolecules. An additional objective is to assemble quantum dots (QDs) and proteins onto biomolecules-functionalized BNNTs surfaces to form novel hybrids with unique properties. The functionalization of BNNTs using water-soluble biomolecules not only endows a good dispersity to BNNTs, but also offers new physiochemical functions to BNNTs.

### 1.3.3.1. Noncovalent Functionalization of BNNTs Using Peptides

A good aqueous dispersion of BNNTs is primarily acquired for biological applications. Peptides have diverse functions due to the diversity of amino acids containing in the sequence of a certain peptide as well as their facilitated conjugating approach. Researchers have suggested that peptides can interact with the hydrophobic sidewall of CNTs via π-π stacking or hydrophobic interactions,[79] which resulted in the individualization of CNTs. In consideration of the similarity between the crystalline structures of BNNTs and CNTs, it was reasonably anticipated that peptides were able to interact with BNNTs in a similar fashion as CNTs. We have employed an aromatic CNTs binding peptide, denoted as B3 (HWSAWWIRSNQS), and inticipated for its interaction with BNNTs via π-π stacking and/or hydrophobic interactions.[75]

In a typical experiment run, BNNTs and peptide solution was mixed, followed by sonication for dispersing, the insoluble materials were removed by centrifugation procedure, then the supernatant above 70-80 % of the dispersion was collected for further characterizations (**Figure 8**). The AFM characterizations on the morphologies suggested that B3-functionalized BNNTs could be excellently dispersed on the mica surface by a slow spin-coat process. The investigations on the conformation suggested that B3 adopted an α-helical conformation in an aqueous solution. B3 showed a considerable dispersing ability for BNNTs. The spectroscopic studies indicated that the π-π interactions existed between B3 and BNNT. The peptide-functionalized BNNTs have promising applications in electronics, optical devices, and biological fields due to excellent dispersion and unique physical properties.

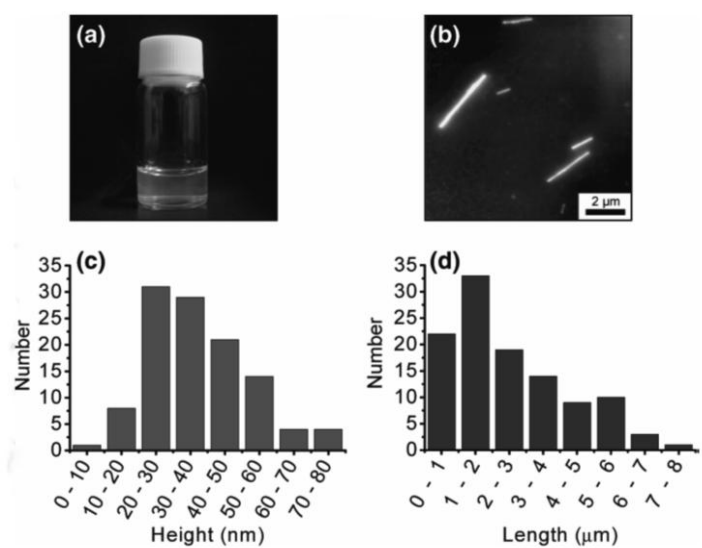

**Figure 8.** (a) Dispersion of B3-BNNT complexes, (b) AFM image of the complexes, and statistical AFM (c) height and (d) length analyses of the complexes collected from 50 different images (the total number of BNNTs was 112). (Reproduced with permission from ref 75. Copyright (2010) American Chemical Society).

### 1.3.3.2. Noncovalent Functionalization of BNNTs Using DNA and Application for Liquid Crystal

In the materials science, the high-concentration dispersion of BNNTs is acquired for increasing their processability, particularly it offers an opportunity to understand the phase behavior in solvents. DNA is a central biomolecule with many amazing properties. The aromatic nucleobases within DNA structures are capable of interacting strongly with the sidewalls of BNNT via π-π stacking and/or hydrophobic interactions (**Figure 9**). Zhi *et al.* have dispersed BNNTs in an aqueous solution containing single-strand DNA (ACG TAC GTA ACG TAC GTA CGT ACG TAC) for obtaining sufficiently high-concentration dispersion.[43] The nematic liquid crystal state of BNNTs was observed from this aqueous high-concentration BNNTs dispersion based upon a simple filtration process. Clearly, an amorphous layer about 5-20 nm was observed in TEM image, suggesting the un-uniform attachment of DNA on the surface of BNNTs. Thermogravimetic analysis (TGA) suggested that 20 wt% DNA were attached on BNNTs surface. Under optimized experimental conditions, a dispersion containing 0.2 wt% BNNTs could be obtained using a single-strand DNA. Apparent shifts were indentified in both absorption and emission spectra of BNNTs and DNA, suggesting that the presence of strong π-π stacking interactions of BNNT and DNA.

The assembly and phase behavior of BNNTs in both solution and bulk states are fundamental for understanding the performance of materials, particularly for liquid crystal. In the aforementioned paper,[43] a BNNTs mat was obtained by treating the filtered mixture of BNNT-DNA at 700 °C for 30 min. The SEM images showed that the mat contains a nematic order phase (**Figure 10**). The ordering behavior depended on the concentration of BNNTs and the filtering process. The functionalization of BNNTs with DNAs may be useful for the development of novel biodevices.

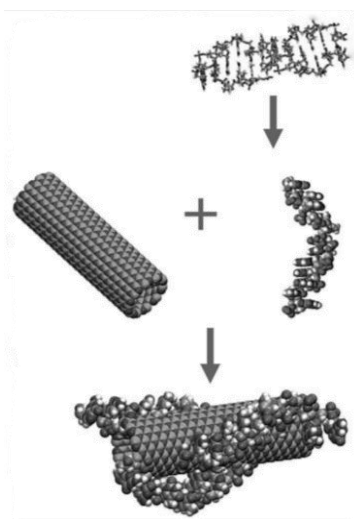

**Figure 9.** The process of fabrication of the DNA-BNNT hybrid. (Reproduced with permission from ref 43. Copyright (2007) John Wily and Son).

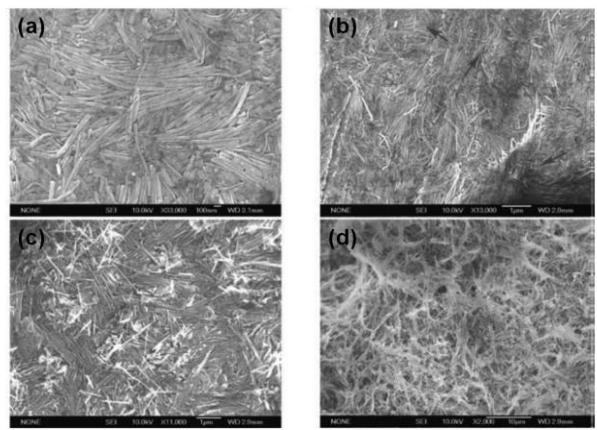

**Figure 10.** (a, b) Nematic ordering of BNNT ensembles (the singularities as shown with arrows in b)); (c) SEM images of the ensembles when a 0.1 wt% BNNTs solution was used for filtering (the BNNTs are less ordered); (d) as the BNNT concentration decreases, the morphology of the DNA–BNNT hybrid mat is affected by the water flow, and no BNNT ordering takes place. (Reproduced with permission from ref 43. (2007) John Wily and Son).

### 1.3.3.3. Noncovalent Functionalization of BNNTs Using Flavin Mononucleotide (FMN) and Application for Visible-Light Emission

Biomolecules, such as peptides and DNA, are able to interact with BNNTs. However, the precise interactions between these biomolecules and BNNTs are much complicated, and are hard to be fully understood at the present stage. The difficulty has led us an idea that small aromatic biomolecules are probably workable on functionalizing BNNTs. Those biomolecules containing aromatic moieties provide simple models for understanding the interactions. Furthermore, unlike macromolecules, small molecules can fully cover the surface of BNNTs, allowing the perfect isolation of BNNTs from their surrounding which is a key point for sustaining their optical properties. FMN is the phosphorylated derivative of vitamin B2. Its molecular structure consists of an aromatic isoalloxazine ring and a phosphate moiety. FMN is anticipated to interact with the hydrophobic surfaces of BNNTs with its isoalloxazine ring via π-π stacking interactions.[76] This establishes a noncovalent sidewall chemical pathway not only for dispersing and functionalizing BNNTs in an aqueous solution, but also for further integrating these innovative BNNTs-based hybrids into new composite materials and devices.

After dispersing the mixture of BNNTs with FMN using a sonicator under a mild condition, the disentanglement of BNNTs in water was achieved and thus FMN-functionalized BNNTs were obtained, denoted as FMN-BNNT nanohybrids (**Figure 11**). The spectroscopic results suggested the presence of the strong π-π interactions between FMN and BNNTs. Optical measurements suggested that FMN-BNNT nanohybrids emitted bright green fluorescence (**Figure 12**). Further studies suggested that the fluorescent intensities were stable in a wide pH and temperature ranges. These nanohybrids are valuable for utilizing as promising visible-light emitters.

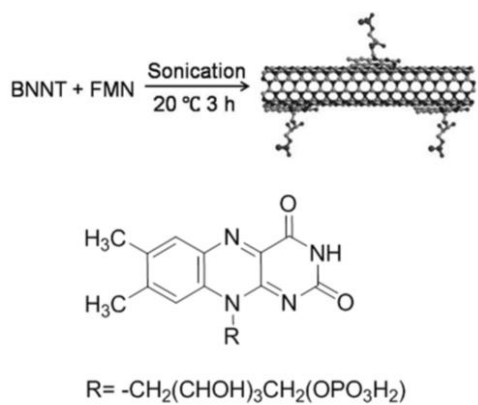

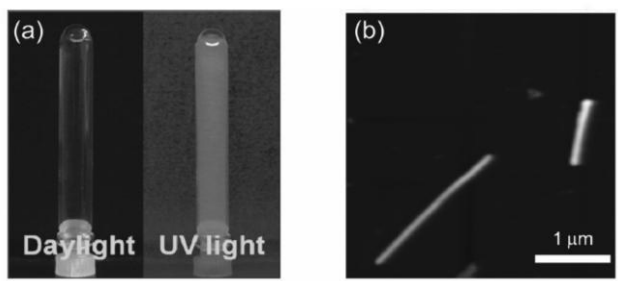

**Figure 11.** The representation of the formation process of a FMN/BNNT nanohybrid. (Reproduced with permission from ref 76. Copyright (2011) American Chemical Society).

**Figure 12.** (a) The dispersion of BNNTs in the aqueous FMN solution under daylight and UV light and (b) the typical AFM image of FMN/BNNT nanohybrid. (Reproduced with permission from ref 76. Copyright (2011) American Chemical Society).

## 1.3.3.4. Noncovalent Functionalization of BNNTs Using Nucleotides and Application for Quantum Dots (QDs) Decoration

The fluorescence of pristine BNNTs locates in the deep-UV range, which hampers the applications of BNNTs in bio-related fields. Nucleotides as the main chemical compounds of DNA and RNA are central molecules in biochemistry, and have structural properties similar to FMN. The idea is that nucleotides can interact with BNNTs via π-π stacking interactions.[77]

Meanwhile, a previous research has shown that guanosine 5′-monophosphate (GMP) caps cadmium sulfide (CdS) QDs via N7 of pyrimidine and/or -NH$_2$ of purine, and P-O-5′-sugar, producing GMP-capped CdS QDs.[80] In this context, free π-electrons should be presented on the GMP-capped QD surfaces. Those π-electrons should be capable of interacting with BNNT sidewalls. Thus, the decoration of BNNTs with GMP-capped CdS QDs is considered as a valuable way for further functionalizing the BNNTs (**Figure 13**). Because these CdS QDs have fluorescence in the visible-light range at longer wavelengths than BNNTs, new fluorescence can be endowed to BNNTs after CdS QDs' decoration.

The used nucleotides included adenosine 5′-monophosphate (AMP), adenosine 5′-diphosphate (ADP), adenosine 5′-triphosphate (ATP), GMP, guanosine 5′-diphosphate (GDP), guanosine 5′-triphosphate (GTP), uridine 5′-monophosphate (UMP), cytidine 5′-monophosphate (CMP), and guanosine (Gua). The potentials of nucleotides for dispersing were compared quantitatively. The results indicated that the order of mononucleotides for better BNNTs dispersion was GMP > AMP ≈ UMP > CMP. The monophosphates were much better than corresponding di- and triphosphates. HR-TEM images showed that the decoration of BNNTs with CdS QDs was achieved by using GMP as a linking reagent (**Figure 14**), which enabled to disperse the largest amounts of BNNTs in aqueous solution among nucleotides. Optical characterizations suggested that CdS/GMP@BNNT hybrids hold a new fluorescence in the visible-light range. The investigations of the interactions between nucleotides and BNNTs were helpful for understanding the chemistry of nucleotide-inorganic interfaces, which are extremely important for developing bio-nanotechnologies. New fluorescence in the visible-light range was endowed to BNNTs after successful decoration with CdS QDs. This will be beneficial in extending the utilization of BNNTs for bio-related applications.

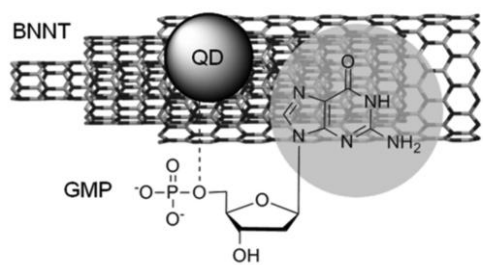

**Figure 13.** Representation of the decoration of BNNT sidewalls with CdS QDs assisted by GMP. (Reproduced with permission from ref 75. Copyright (2011) Royal Society of Chemistry).

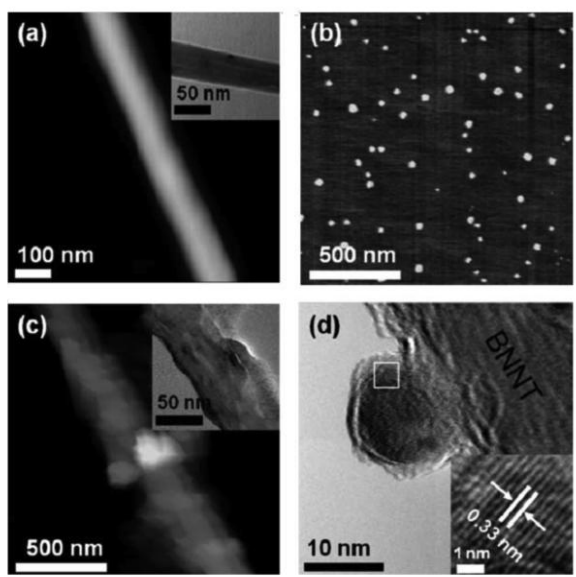

**Figure 14.** Representation of the decoration of BNNT sidewalls with CdS QDs assisted by GMP. (a) High-resolution AFM image of a GMP-modified BNNT. The inset shows the corresponding TEM image. (b) AFM image of GMPcapped CdS QDs. (c) High-resolution AFM image of CdS/GMP@BNNT hybrids. The inset shows the corresponding TEM image. (d) Representative HR-TEM image of the hybrids. The inset shows the enlarged square portion. (Reproduced with permission from ref 75. Copyright (2011) Royal Society of Chemistry).

### 1.3.3.5. Noncovalent Functionalization of BNNTs Using Polysaccharides and Application for Proteins Decoration

The previous investigations showed that peptides, DNA, FMN, and nucleotides were able to work on the sidewall functionalization of BNNTs. However, these molecules are frequently produced by precise synthesis. To promote the prospective applications, new approaches for sidewall functionalization with lower-cost, better efficiency, and biocompatibility are highly demanded. Reasonably, natural polysaccharides are excellent candidates due to their cheapness and biocompatibility. Among them, gum arabic (GA) is one of the abundant polysaccharide in nature with highly branched complex molecular properties. GA is also highly water-soluble and easily chemical post-functionalization. Just recently, the use of GA has started to be extended into nanoscience and nanotechnology. We have shown that GA can be used to disperse and functionalize BNNTs in an aqueous solution.[78] The disentanglement of BNNT from raw bundled materials was achieved by functionalizing them with GA. These GA-functionalized BNNTs offered an opportunity for studying the physiochemical properties of single disentangled BNNTs. Afterwards, to show an advantage of GA-functionalized BNNTs, several functional proteins (streptavidin (SAv), lysozyme (Lyz), bovine serum albumin (BSA), and immunoglobulin G (IgG)) were successfully decorated onto the GA-functionalized BNNT surfaces by using the electrostatic interactions between proteins and GA-functionalized BNNTs (**Figure 15, 16**). The successful assembly of proteins on the surface of GA-functionalized BNNTs shows an importantly initial step for developing BNNT-based bio-devices.

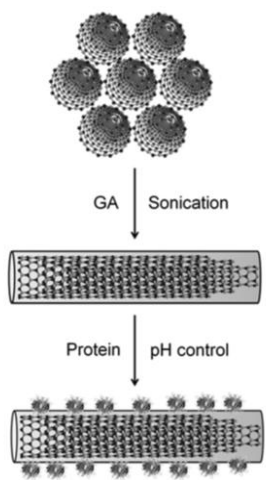

**Figure 15.** Schematic representation of disentanglement of BNNTs via functionalization with GA for proteins immobilization. (Reproduced with permission from ref 78. Copyright (2012) Royal Society of Chemistry).

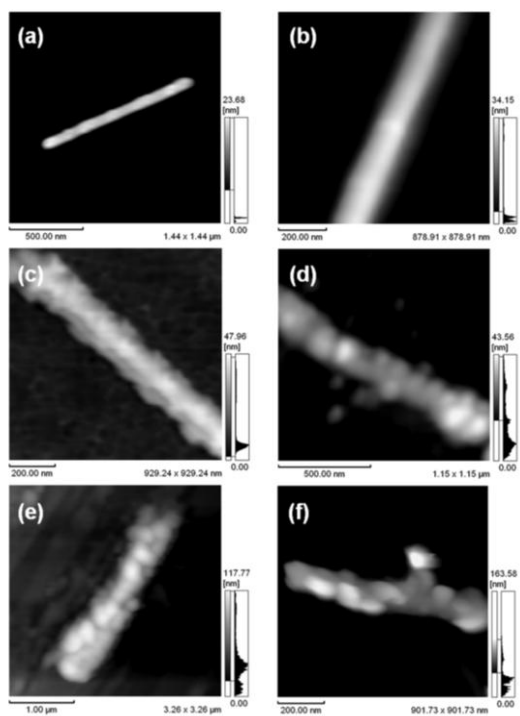

**Figure 16.** (a, b) AFM images of GA-functionalized BNNTs; (c) SAv (d) BSA (d) Lyz and (e) IgG on GA-functionalized BNNTs. (Reproduced with permission from ref 78. Copyright (2012) Royal Society of Chemistry).

### 1.3.3.6. Noncovalent Functionalization of BNNTs Using Lipids and Application for Length Cutting

The great promise of BNNTs for many biomedical applications has been realized recently. For example, BNNTs offer high potentials for the use of boron neutron capture therapy[66] due to the high neuron catching ability of boron. However, an aqueous dispersion of BNNTs must be achieved prior to any biological experiments. Meantime, CNTs with short lengths are more biocompatible and less toxicity in the tested cells and animals than long ones.[81] The original lengths of as-produced BNNTs can be up to several tens of μm, which is too long for testing in most biological samples, particularly for the internalization into living cells.[68] The toxicity of BNNTs on human embryonic kidney (HEK293) cells has recently found.[68] The origin of the toxicity is probably due to the large axial size. It is therefore necessary to develop a procedure for cutting and controlling the lengths. Very recently, Lee *et al.* have functionalized BNNTs with a PEGylated phospholipid [methoxy-poly(ethylene glycol)-1,2-distearoyl-sn-glycero-3-phosphoethanolamine-N conjugates (mPEG-DSPE)] (**Figure 17**).[82] The fatty acid chains of mPEG-DSPE can strongly bind around the sidewall of BNNTs through van der Waals, charge transfer, and hydrophobic interactions, whereas mPEG chain is interacting with water molecules through hydrogen bonds, thereby leading to mPEG-functionalized BNNTs greatly dispersible in water. No notable precipitations were found in the water-dispersion of BNNTs containing mPEG-DSPE after keeping up to three months. This is a great improvement in the stability of dispersed BNNTs in water. The hydrogen bonds between water molecules and hydrophilic tail of mPEG are found to be essential for keeping the BNNTs dispersion stable. Furthermore, the length distribution can be controlled within the range of 500 nm by simply prolonging the sonication time up to 20 h. This procedure is easy to control and scale up. Although the exact function of mPEG molecules in the cutting process is required to be understood more carefully. This finding is still very useful for integrating BNNTs into valuable biological systems.

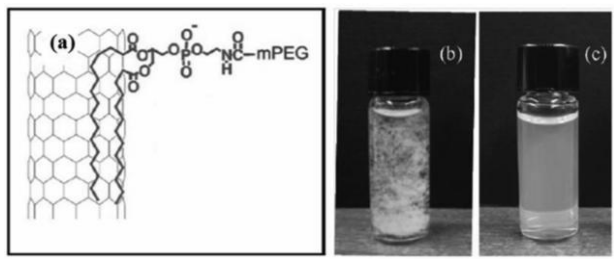

**Figure 17.** (a) Schematic representative of a BNNT functionalized by an mPEG-DSPE molecule. (b) The extracted BNNT bundles in ethanol. (c) Well-dispered mPEG-DSPE/BNNTs in water (after 2 h of sonication). (Reproduced with permission from ref 82. Copyright (2012) American Chemical Society).

## 1.4 Summary and Perspective

In this contribution, the latest advancements in noncovalent functionalization of BNNTs with polymers and biomolecules were outlined. The noncovalent interactions between BNNTs and functional molecules allow their efficient attachment onto the sidewall of BNNTs, thereby endowing a good dispersity and new chemical functions to BNNTs. Meanwhile, the application of those functionalized BNNTs for superhydrophobic surface, visible-light emission, QDs decoration, protein immobilization, and length cutting have been further summarized herein. This contribution should be able to promote the interests in the functionalization of BNNTs using polymers and biomolecules. Although several essential progresses have been achieved, the application of BNNTs is still far away from the practice. Many issues are still not fully understood so far. For example, the toxicity and biocompatibility of these functionalized BNNTs are needed to be fully investigated *in vitro* and *in vivo*. The cell-specific recognition function is also acquired to be endowed to BNNTs surface for internalizing them into cells in the future.

To the end, great advancements have been recently made in the development of 2-D layered nanomaterials such as graphene,[83,84] graphene oxide,[85] BN nanosheet,[86] and nanoribbons.[87] Well-isolated 2-D layered nanomaterials can be considered as just surface-like materials, because the surface and its edge structures are particularly dominant for the properties of those materials.[88] In other words, the physiochemical properties of these novel nanomaterials are extremely sensitive to the surface species and the surrounding, thereby it is important to modify their surfaces and edges.[88,89] With consideration of similarity of the chemical structures, it is possible to functionalize the surface and its edge of those 2-D nanomaterials using various polymers and biomolecules. This should be a new research domain of nanoscience and nanotechnology in the coming years.

**Corresponding Author**

* To whom correspondence should be addressed. E-mail addresses: zhenghong.gao@u-bordeaux1.fr

**Author Contributions**

All authors discuss the contents of the manuscript. Z. G. writes the draft. Z. G. and T. S. modify the manuscript. All authors have given approval to the final version of the manuscript.

**ACKNOWLEDGMENT**

Z.G. thanks Dr. T. Sawada and Mr. K. Fujioka for his kind assistance in many experimental measurements. Z.G. and T.S. thank Prof. M. Komiyama and Prof. M. Aizawa for helpful discussion during the preparation of the experiments and this manuscript.